# Defect studies in strain-relaxed $Si_{1-x}Ge_x$ alloys

Şeref KALEM

TÜBITAK – BILGEM – UEKAE National Research Institute of Electronics and Cryptology,

Gebze 41470 Kocaeli, Turkey

seref.kalem@tubitak.gov.tr

Tel:+90 262 648 1622

# Abstract

Raman light scattering, low temperature photoluminescence, light scattering tomography and hydrogenation were used to investigate optical properties of defects in strain-relaxed $Si_{1-x}Ge_x$ ( $0.05 \leq x \leq 0.50$ ) alloys. The photoluminescence emission was characterized by typical zero-phonon, phonon-assisted and dislocations related emissions, which are dependent on Ge composition x. However, luminescence spectra exhibit above band-gap features which are likely associated with the presence of Si-rich regions in the alloys. The results are correlated with light scattering tomography revealing the presence of dislocations and Si precipitates. The excess peak at 519 $cm^{-1}$ at Ge-rich samples is supportive of this observation. At low Ge content, a dislocation related band ( D2 line ) at 14204 Å dominates D-bands emission for x < 0.25 while overall D-band emission intensity decreases with x. Hydrogenation was found to enhance D-band emission indicating a passivation of non-radiative recombination centers inside dislocation cores. Si-Si, Si-Ge and Ge-Ge phonons ( TO, TA and LA ) which are participating in luminescence emission evolve with increasing Ge content and Ge-Ge and Si-Ge TO lines dominate the Raman spectrum to the detriment of the Si-Si TO phonon line. Raman spectra reveals the presence of alloy fluctuations and possible presence of Ge particles particularly in Ge-rich samples.

**Key Words:** $Si_{1-x}Ge_x$ alloys, photoluminescence, Raman light scattering

**PACS :** 78.20.-e, 78.30.-j, 78.55.-m

1. **Introduction**

Defects in strain-relaxed $Si_{1-x}Ge_x$ alloy layers grown by epitaxy on Si substrates consist of growth and process induced defects [1, 2]. Due to lattice mismatch of 4% between Si and Ge, relatively thick layers are relaxed into defects through the creation of misfit dislocations and resulting threading dislocations running to the surface. The misfit dislocation structure present in compound semiconductors and their association with photoluminescence emission have been reported earlier [3]. These emissions were observed as D-band dislocation luminescence and have been associated with interstitial type defects as supported also by DLTS measurements [4].

There has been an extensive study of dislocation-related photoluminescence in Si [5-7] and $Si_{1-x}Ge_x$ alloys [8-10]. From experimental point of view, D1 / D2 and D3 / D4 bands have similar origin due to similarities in their optical properties. The radiative recombination process in these alloys can originate either from impurity related transitions or from the intrinsic property of the dislocation [6].

The effect of hydrogenation on dislocations in relaxed $Si_{1-x}Ge_x$ is to enhance the near band-gap emission, which was correlated with a quenching in the D-line luminescence especially for D3-D4 bands [11]. It was argued that the hydrogen passivates non-radiative centers inside dislocation cores. Another study has found that a strain-relaxed $Si_{1-x}Ge_x$/Si interface is effective in trapping hydrogen during plasma treatment. Long micro cracks were observed at the interface due to trapping of indiffused hydrogen [12]. Hydrogenation of Si was found to be effective in passivating D1/D4 bands while enhancing D2 and D3 emissions [5].

## 2. Experimental

$Si_{1-x}Ge_x$ alloy structure consists of a buffer layer which was compositionally graded with a Ge content increase rate of 10% / μm which was followed by the $Si_{1-x}Ge_x$ layer of 4 μm thickness ( Boron doped to $p = 5 \times 10^{15} cm^{-3}$ and an Sb doped ( $n = 10^{19} cm^{-3}$ ) $Si_{1-x}Ge_x$ cap layer of 0.3 μm. The structure was grown by molecular beam epitaxy on p-type Si ( 100 ) doped by B to $p = 1 \times 10^{19} cm^{-3}$. Raman spectra were collected at room temperature in a backscattering geometry using a frequency doubled Nd:YAG laser 532 nm line as the excitation source. The photoluminescence was excited by either an $Ar^+$ laser 514 nm line or a Krypton laser 647 nm line of 150 mW. The signal was dispersed using a 1m Spex monochromator and detected by a liquid nitrogen cooled Ge detector in the spectral range 1.0 - 1.8 μm. The samples were mounted in a liquid He bath cryostat at 4.2 K. Laser scattering tomography (LST) as an effective investigation method of precipitates and dislocations used a Nd:YAG laser at 1060 nm with confocal imaging at Brewster angle mode with a penetration depth of 5000 nm.

## 3. Results

### 3.1 Photoluminescence
### A - Band-edge emission

There is a smooth variation of the characteristic photoluminescence ( PL ) features in the strain relaxed SiGe alloys with increasing composition x. Figure 1 shows examples of the low temperature near band-gap PL exhibiting zero-phonon ( $X^{NP}$ ) and phonon (see section 4.1 Raman results for phonon energies ) assisted emissions. The PL was excited using Kr laser line of 647 nm and the measurement was taken at 4.2 K. In order to distinguish between the band gap energies, the calculated band gaps $E_g$ (x) using the analytical expression [8] are also indicated by arrows in the figure. In the figure, we observe a significant contribution from the substrate due to relatively long laser penetration depth. The variation of the PL peak energy at band edge is clearly evident when we used Ar ion laser as indicated in Figure 2. The change of the peak energy as deduced from Figure 2 is shown in Table 1. Also, included in this work are the values for the excitonic band-gap of SiGe alloys and corresponding corrected compositions as deduced from the PL data using the expression reported earlier by Weber [8, 13].

### B-Dislocation luminescence

Dislocation related PL emission ( D-lines ) is shown in Figure 3 as a function of composition x and the frequencies of these lines are summarized in Table 2. For Si rich samples there are four distinct peaks which are located at 15365Å, 14204 Å, 13348 Å and 12513 Å which belong to D1, D2, D3 and D4 emissions, respectively. It was found that both positions and intensities of the D-line emissions change with alloy composition x. Particularly, in $Ar^+$ laser excitation, there is a considerable shift to low energy for alloys with composition greater than x = 0.25. These changes are summarized in Figure 4 and Figure 5 as a function of alloy composition. As shown in the Figure 4, the separation between the lines D1 and D2 is decreasing with x, which was followed by similar behavior in the emission intensity as indicated in Figure 5. The PL intensity of D1 line for Si-rich alloys is much stronger than the other D-lines and peaks at around x = 0.20. The PL intensity for all the D-lines decreases in Ge-rich alloys. Another important feature with D-lines is the asymmetric lineshape of the D2 emission for compositions greater than x = 0.35. Particularly, for the sample with x ≥ 0.50, we observe a significant shoulder at the low energy side of this peak, that is 14588 Å.

### 3.3 Hydrogenation

The $Si_{1-x}Ge_x$ alloys have been hydrogenated in a hydrogen plasma reactor at 200 $^o$C for 1 hour. The effect of hydrogenation is a slight decrease in the band-edge luminescence and an enhancement in PL intensity of the D-lines as shown with a typical example in Figure 6 for a sample containing 35% Ge. Note that the figure shows both the band-edge and the D-line emissions. The hydrogenation was found to be the most effective on the D3 emission, where the PL intensity is increased by more than 45%. We assume that the hydrogenation decreases the activity of non-radiative recombination centers at dislocations resulting in relatively enhanced PL at D-lines.

## 4. Discussion
### 4.1 Raman light scattering

Figure 7 shows the unpolarized room-temperature Raman spectra of the strain-relaxed single crystal $Si_{1-x}Ge_x$ alloys ( x = 0.05, 0.15, 0.25, 0.35, 0.50 ) which were grown by MBE on p-type Si (100). Peak positions of these alloys are given in Table 3 for several compositions x. In Figure 7, each spectrum is characterized by three dominant peaks centered at around 500 cm$^{-1}$, 405 cm$^{-1}$ and 290 cm$^{-1}$, which correspond to scattering of light involving TO phonons of Si-Si, Si-Ge and Ge-Ge stetching vibrations, respectively. The origin of these peaks and their dependence on composition x have already been reported by several groups [13-15]. In addition to these peaks, each spectrum contains two weak peaks (shoulders) between Si-Si and Si-Ge modes. There is another weak peak at 572 cm-1 which grows in intensity with increasing composition x. This band is attributed to the second order phonon mode of Ge-Ge vibrations. The last peak which exists as a weak shoulder to main peak for x = 0.25 and 0.35 can be clearly distinguished at 519 cm$^{-1}$ when the main Si-Si mode moves toward low frequencies for x = 0.50. As shown in Figure 7, the weak peaks S1 at 487 cm$^{-1}$ and S2 at 436 cm$^{-1}$ for x = 0.05 exhibit similar behavior to Si-Si and Si-Ge peaks with composition change, respectively. The Raman shift of these weak peaks are compared to Si-Si and Si-Ge modes in Fig.8. In the spectra, we observe a deviation from the ideal probability of occurrence (( 1-x )$^2$, 2x ( 1-x ), x$^2$ ) for intensity distribution of Si-Si, Si-Ge and Ge-Ge bands, respectively. For x = 0.50, this difference is partly due to the error in x value which was corrected as 0.45 from PL measurements. Also, the filter used in the measurement caused a decrease in the observation of Ge-Ge band intensity.

**4.2 Above band-gap emission**

For the alloys having compositions larger than x = 0.15, we observe a PL emission bands above the excitonic band-gap of the corresponding SiGe alloy. These above gap emissions are typical of bulk Si peaks and therefore it can be attributed to the substrate or the separation of the alloy into separate phases, one of which is "pure" Si. However, we can not rule out of the possibility of having an exciton diffusion and the presence of possible Silicon rich regions inside the alloy. It has already been shown that the FE's in Si can have diffusion length greater than 20 µm at low temperatures [16]. But, we dont rule out the fact that the laser light can activate the Si precipitates directly. Si-rich regions in our SiGe alloys can indeed explain the observed above gap features. In the Raman spectra, the residual weak peak at 519 cm$^{-1}$ can be the evidence of such Si-rich regions. This is also supported by laser scattering tomography measurements as shown in Figure 9 where we observe particles(precipitates) which are likely the image of such regions. Excitation energy dependence of the PL can give further clue about the origin of the above gap emission lines. Figure 10 shows spectra taken on the sample containing 50% Ge using

Ar$^+$ ion laser ( 514 nm ) and Kr laser line of 647 nm. At 514 nm, the optical penetration depth 1 / 2α is about 0.3 μm for 25% Ge containing alloy [17]. The optical penetration depth decreases sharply with increasing Ge content, thus the Ar$^+$ laser line is almost totally absorbed within the cap layer of Si$_{1-x}$Ge$_x$ ( 0.3 μm and Sb doped to $10^{19}$ cm$^{-1}$). Figure 10 shows very weak above gap emissions at 11360 Å ( X$^{NP}$ of x = 0.15) and at 11860 Å ( X$^{NP}$ of x = 0.35 ) for 50 % alloy and an above gap emission at 11121 Å ( X$^{NP}$ of x = 0.05 ) for 25% alloy . The weak emissions could be due to relatively small penetration depth. The penetration depth for Krypton ( 647 nm ) can be as thick as the layer for Si-rich samples but should be much shorter in Ge-rich samples due to stronger absorption in Ge. Note that the overall layer thickness for x = 0.50 alloy is about 9 μm, that is two times thicker as compared to x = 0.05 alloy ( 4.8 μm ).

**4.3 Defects related to dislocations**

From the energy separation of D1 and D2, one can assign D1 to the O$^\Gamma$ for Si-rich alloys ( 65 meV ) and TO phonon replica of D2 ( 59 meV ), respectively. The separation of D3 and D4 bands is identical to the separation of the BE$^{NP}$ and BE$^{TO}$ lines, thus suggesting that D3 is the TO-phonon replica of D4 as it is well known in pure Si. The D-band luminescence exhibits different properties depending on excitation energy and the Ge content in the alloy. With Ar$^+$ laser excitation, all the D-bands shift to lower energies and the separation between D1 and D2 decrease with x. The change could be due to heating related effects as a result of increasing dislocation density with x. With Kr 647 nm line excitation, D-band positions seem to be independent of composition x. This difference can be attributed to the fact that Ar$^+$ laser probes only the cap layer while the most of the PL contribution comes from the bulk region with Krypton laser. Particularly for x = 0.50 alloy, band edge PL is the dominant feature with Ar$^+$ line, while D-bands emission dominates the spectra with 647nm line excitation. This suggests that D1 and D2 emissions originate from the region close to the interface as shown in Figure 10. Further evidence comes from the 514 nm excited D-band PL in x = 0.35 and x = 0.50 alloys where the emission intensity is very weak as shown in Figure 11. The weak intensity can be attributed to an increased non-radiative recombination since dislocation density enhances at high Ge concentrations. Surface states or Sb related defects can play an important role in this type of recombination. Also, it is found that for these compositions D-band emissions are shifted to low energies. For both excitation energies, D-band emission intensity decreases suggesting an increase in non-radiative recombination.

**Alloy fluctuation**

Weak bands observed between Si-Si and Si-Ge modes in Raman spectra in Figure 7 can be explained by a model using 216-atom cubic supercell [13]. According to this model, these weak features are due to localized Si-Si vibrations in the neighborhood of one or more Ge atom. The large mass of Ge is to decrease the force constant, thus pulling modes out of the pure Si LO-TO band to lower frequencies. The eigen-frequency $\omega$ is then determined by the effective stretching force constant $k_i$ for each band and the mass of Si, $m_{Si}$. In the case where Si has 2 neighbors ( $Si_2$-Si-$Ge_2$ ), $\omega$ would be given by [13]:

$$\omega^2 = (1/m_{Si}) \sum k_i = (k_0/3m_{Si})(2+2+1+1)$$

so that $\omega/\omega_0 = 0{,}87$, thus the frequency $\omega_0 = 520$ cm$^{-1}$ in pure c-Si should be reduced by a factor of 0.87 to 450 cm$^{-1}$, which is in a very good agreement with our observation. From these considerations, we assign two other shoulders at 475 and 430 cm$^{-1}$ in the Raman spectra to $Si_3$-

Si-Ge$_1$ and Si$_1$-Si-Ge$_3$ bonds, respectively.  These are the Raman modes having relatively weak intensities which are unlikely to be observed by light scattering tomography.

## 5. Conclusion

We have investigated in a systematic way the optical properties of defects in strain-relaxed alloys ( $0.05 < x < 0.50$ ) using Raman scattering, photoluminescence, laser scattering tomography and hydrogenation measurements.  It is found that Si$_{1-x}$Ge$_x$ layers exhibit above band gap emission indicating the presence of Si rich regions in the layers.  This is supported by by laser scattering tomography and the excess peak observed at 519 cm$^{-1}$ in the Raman spectra.  The intensity and energetic positions of D-band luminescence depend on Ge composition and excitation energy.  Sb probably is involved in D-line shift in Ge-rich samples. Hydrogenation at 200 $^o$C for one hour passivates non-radiative recombination centers inside dislocation cores.  The results indicate a alloy fluctuation and possibly the presence of Ge particles in Ge-rich alloys.

## Acknowledgement

This work was supported by EU-FP6-2002-IST-1 CADRES project #506962.  The author would like to thank G. Kissinger, IHP; A.Nylandsted-Larsen, University of Aarhus; E.Lavrov and J. Weber, TU Dresden for providing samples and help with experiments and fruitful discussions.

**REFERENCES**

[ 1 ]  A. Nylandsted-Larsen, *Materials Science and Engineering,* **B, 71**, (2000), 6

[ 2 ]  E.V. Monakhov, S.Yu. Shiryaev, A. Nylandsted-Larsen, J. Hartung, G. Davies, *Thin Solid Films,* **294**, (1997), 43.

[ 3 ]  J. Weber, *Solid State Phenomena,* **37-38**, (1994), 13.

[ 4 ]  P. N. Grillot, S. A. Ringel, J. Michel, E. A. Fitzgerald, *J.Appl.Phys*., **80**, (1996), 2823.

[ 5 ]  R. Sauer, J. Weber, J. Stolz, E. R. Weber, K. H. Küsters, H. Alexander, *Appl. Phys. A,* **36**,(1985),1.

[ 6]  K. Weronek, J. Weber and H. J. Queisser, *Phys.Stat.Sol.(a),* **137**, (1993), 543

[ 7]  W. Staiger, G. Pfeiffer, K. Weronek, A. Höpner, J. Weber, *Materials Science Forum,* **143-147**, (1994), 1571.

[ 8]  J. Weber and M. I. Alonso, *Physical Review B,* **40**,(1989), 5683.

[9 ]  S. A. Shevchenko and A. N. Izotov, *Phys.Stat.Sol.(c),* **2**, (2005), 1827.

[10]  E. A. Steinman, V. I. Vdovin, T. G. Yugova, V. S. Avrutin and N. F. Izyumskaya, *Semicond.Sci.Techn.* **14**, (1999), 582.

[11]  A. Daami, G. Bremond, M. Caymax and J. Poortmans, *J.Vac.Sci. Techn.B,* **16**, (1998),1737.

[12]  P. Chen, P. K. Chu, T. Höchbauer, M. Nastasi, D. Buca, S. Mantl, N. D. Theodore, T. L. Alford, J. W. Mayer, R. Loo, M. Caymax, M. Cai, S. S. Lau, *Applied Physics Letters,* **85**, (2004), 4944.

[13]  M. I. Alonso and K. Winer, *Physical Review B,* **39**, (1989), 10056.

[14 ]  A. Renucci, J. B. Renucci and M. Cardona, in *Light Scattering in Solids*, edited by M. Balkanski (Flammarion, Paris, 1971), 326.

[15] G. M. Zinger, I. P. Ipatova and A. V. Subashiev, *Sov.Phys.Semicond.*, **11**, (1977), 383.

[16] Y. H. Chen and S. A. Lyon, *IEEE J. of Quantum Electronics*, **25**, (1989), 1053.

[17] M. Holtz, W. M. Duncan, S. Zollner, R. Liu, *J. Appl. Phys.*, **88**, (2000), 2523.

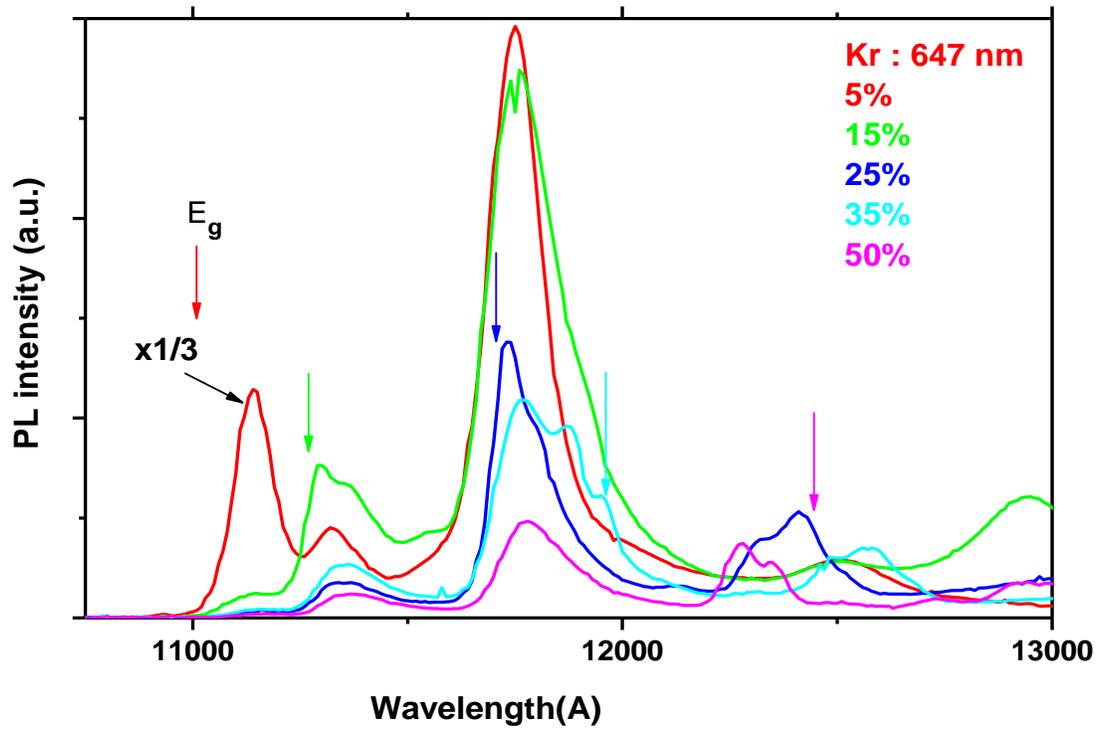

**Figure 1.** Near-band-gap photoluminescence spectra for several $Si_{1-x}Ge_x$ alloys. The spectra were taken at 4.2 K using a Krypton laser 647 nm line. The calculated band gaps $E_g$ (x) are indicated by arrow marks.

**Table 1**

Excitonic band-gaps deduced from the photoluminescence using Ar$^+$ laser line of 514 nm, layer thickness and the composition values which were corrected from PL data. The increasing thickness with **x** is due to the compositional grading of the buffer at a rate of 10% / μm.

| Ge content, x | Thickness (μm) | X$^{NP}$ wavelength (Å) | x (corrected) |
|---|---|---|---|
| 0.05 | 4.8 | 11130 | 0.08 |
| 0.15 | 5.8 | 11286 | 0.14 |
| 0.25 | 6.8 | 11730 | 0.26 |
| 0.35 | 7.8 | 11873 | 0.33 |
| 0.50 | 8.8 | 12360 | 0.45 |

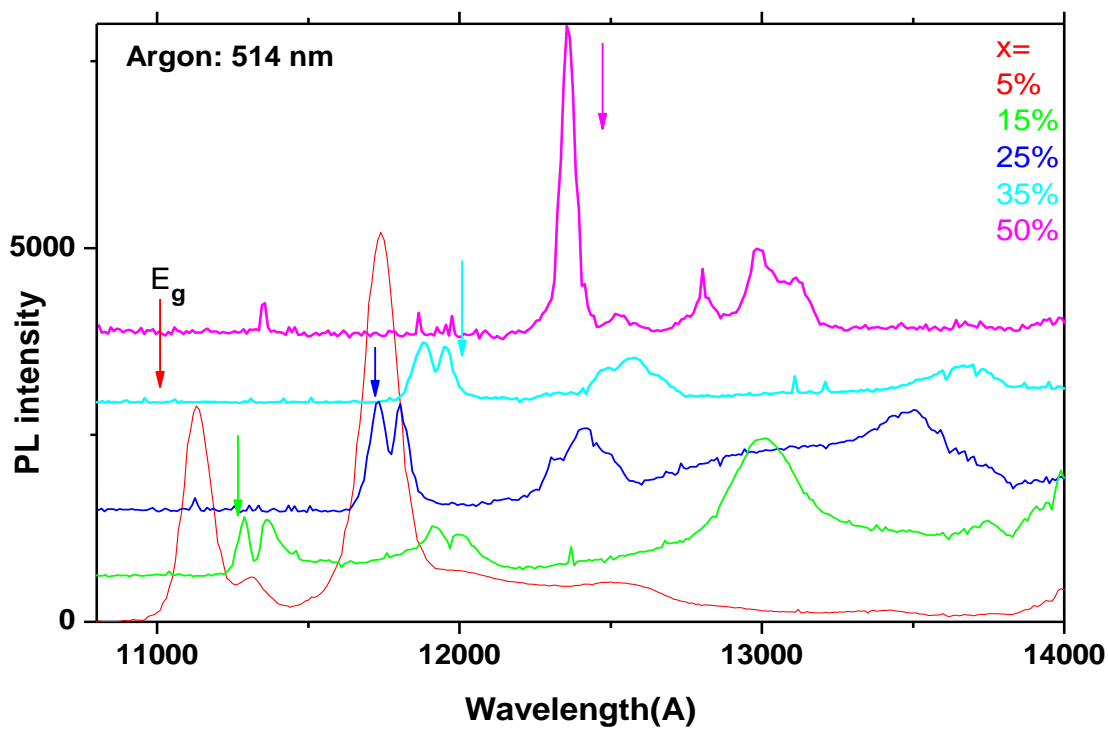

**Figure 2.** 4.2 K Ar$^+$ laser 514.5 nm line excited PL spectra of Si$_{1-x}$Ge$_x$ alloys as a function of Ge content. The calculated band gaps **E$_g$** (x) are indicated by arrow marks.

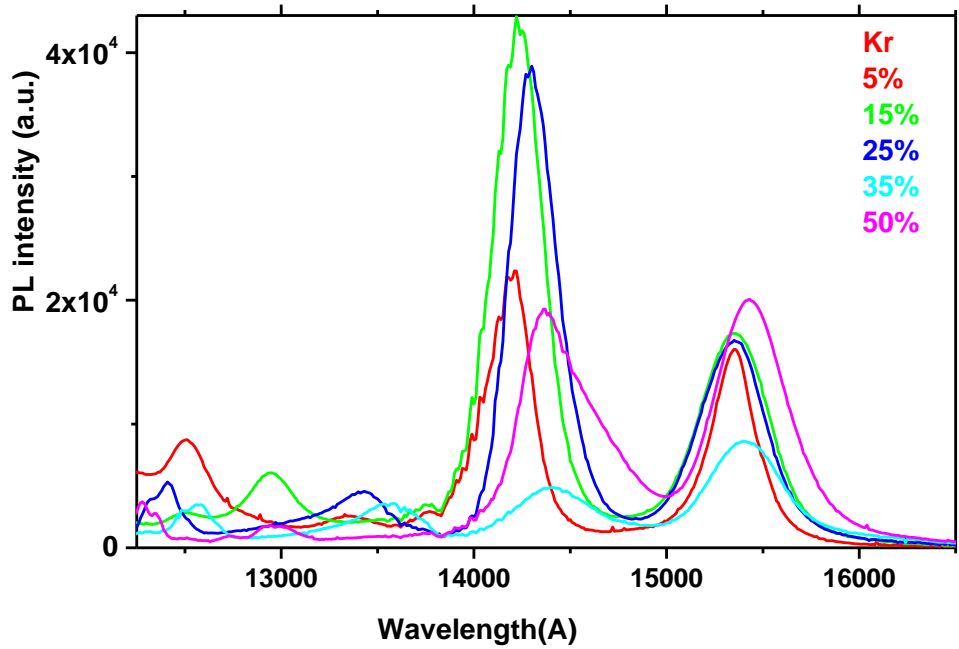

**Figure 3.** D-band photoluminescence in $Si_{1-x}Ge_x$ alloys versus Ge content x ( 4.2 K and 647 nm excitation ).

**Table 2**

D-band peak positions for several alloy compositions as deduced from Figure 3.

| Ge content, x | D1 (Å) | D2 (Å) | D3 (Å) | D4 (Å) |
|---|---|---|---|---|
| 0.05 | 15365 | 14204 | 13348 | 12513 |
| 0.15 | 15365 | 14236 | 12944 | 12500 |
| 0.25 | 15346 | 14286 | 13405 | 12412 |
| 0.35 | 15404 | 14402 | 13522 | 12576 |
| 0.50 | 15423 | 14368 | 12971 | 12731 |

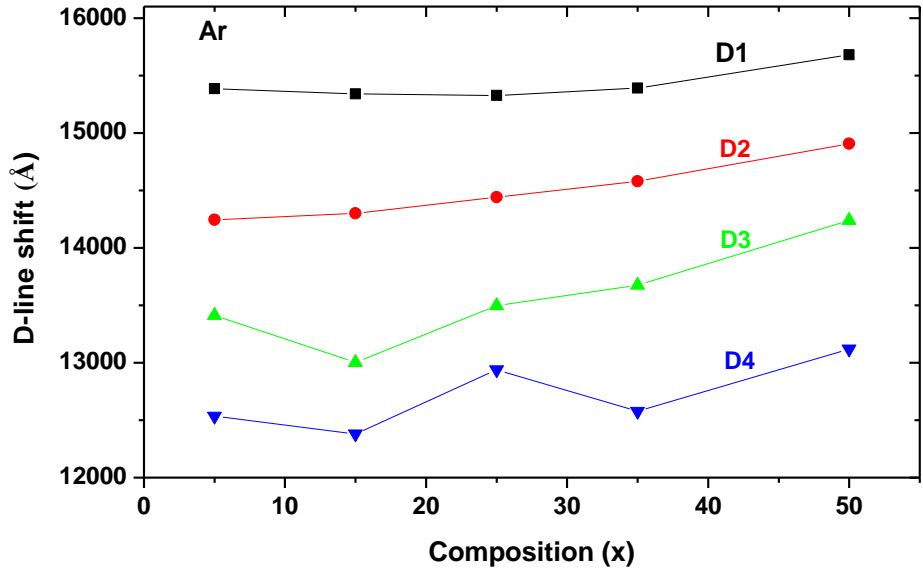

**Figure 4.** Position change in D-band luminescence lines with x. The spectra were taken at 4.2 K and excited with Ar$^+$ laser 514.5 nm line.

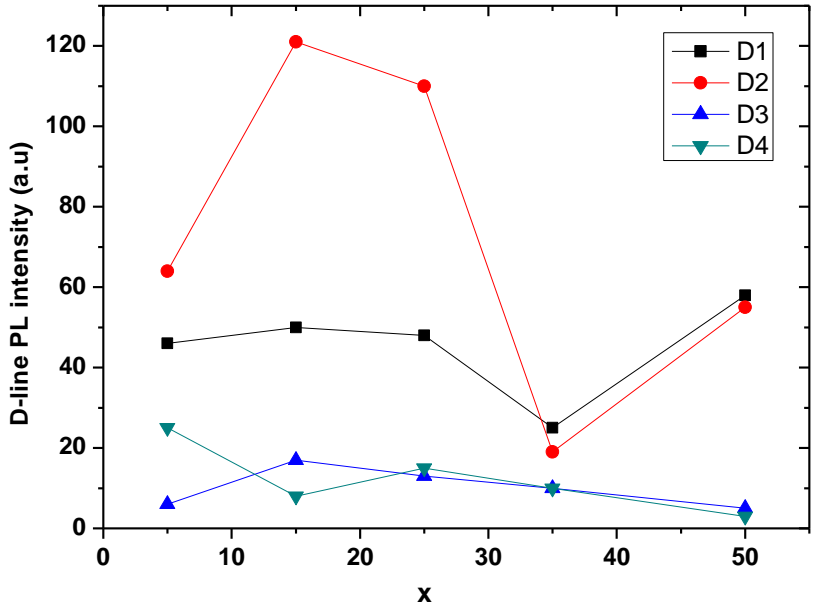

**Figure 5.** The PL intensity evolution of D-bands as a function of Ge content.

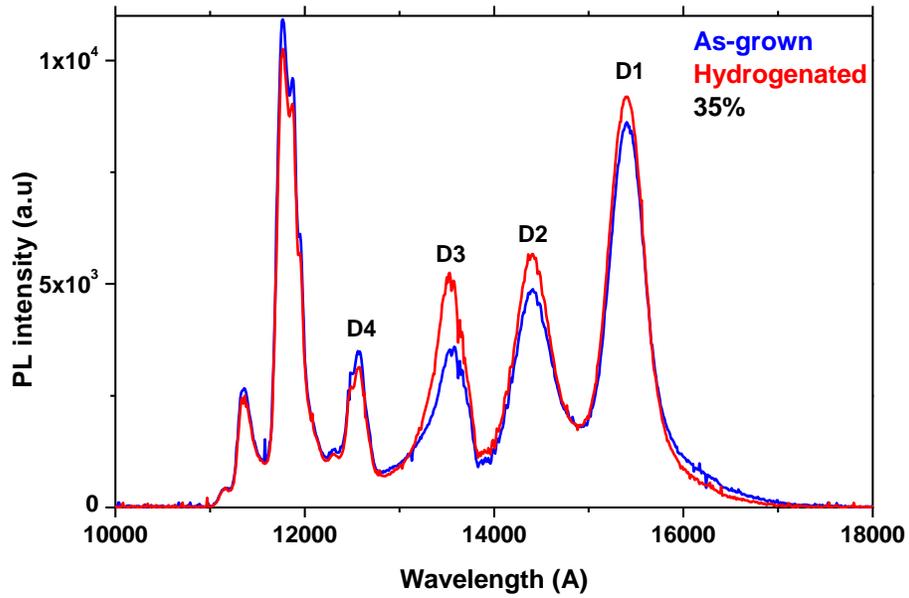

**Figure 6.** Photoluminescence spectra of plasma treated and as-grown $Si_{1-x}Ge_x$ alloy (x = 0.35). The samples were excited at 4.2 K by 647 nm line of Krypton laser. The figure shows both band-edge and D-line emissions.

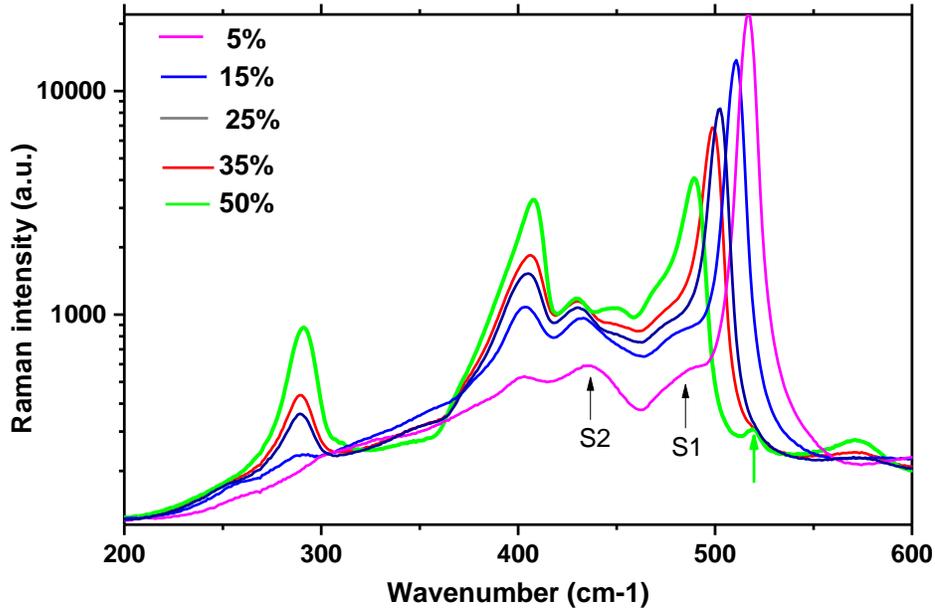

**Figure 7.** Evolution of Raman spectra of SiGe alloys deposited on p-type Si (100) as a function of Ge composition, x. These spectra were obtained at room temperature using a Nd:YAG laser 532 nm line. The arrow indicates residual Si-Si mode in the sample having Ge concentration of 50%.

**Table 3**

Raman peak frequencies in cm$^{-1}$ for the main Raman active modes for SiGe alloys for 5 compositions. Also shown are the weak shoulders (S1 and S2) between Si-Si and Si-Ge peaks.

| x | Si-Si | S1 | S2 | Si-Ge | Ge-Ge |
|---|---|---|---|---|---|
| 0.05 | 517 | 487 | 436 | 402 | |
| 0.15 | 511 | 479 | 432 | 403 | 289 |
| 0.25 | 502 | 476 | 431 | 405 | 289 |
| 0.35 | 499 | 474 | 430 | 406 | 289 |
| 0.50 | 490 | 469 | 429 | 408 | 291 |

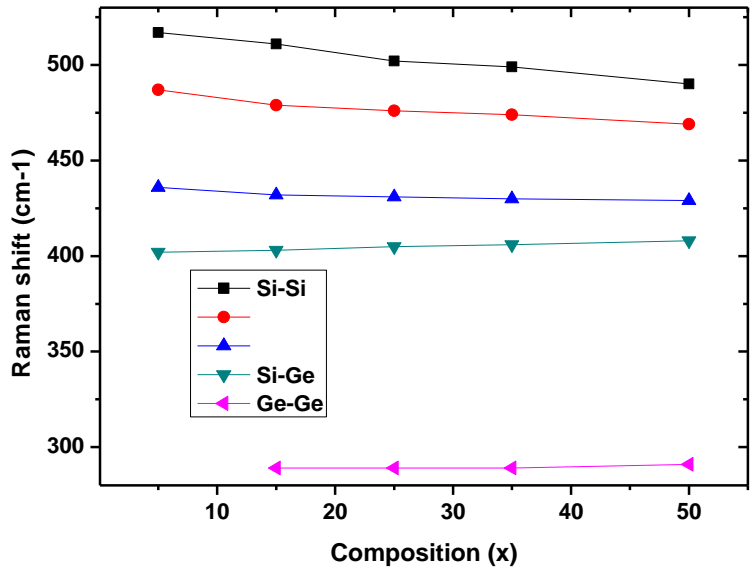

**Figure 8.** Raman shift of the weak peaks and major phonon bands ( Si-Si, Si-Ge and Ge-Ge ) in SiGe alloys as a function of Ge content.

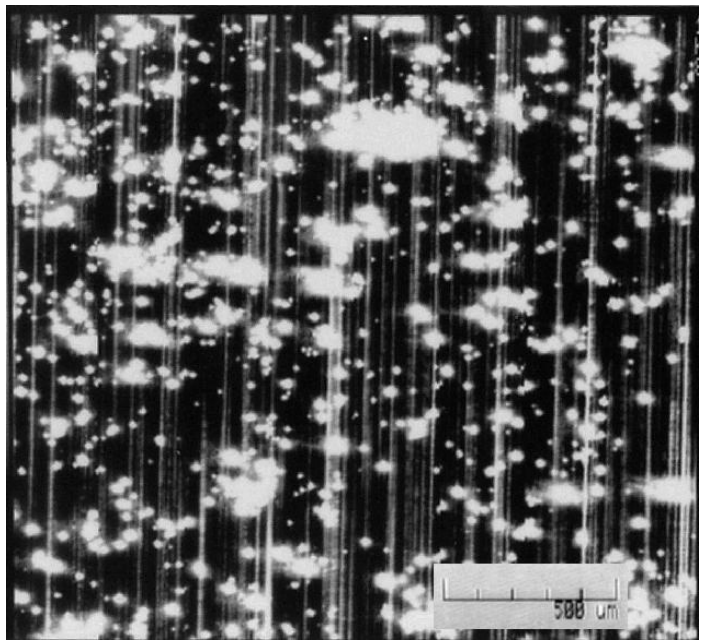

**Figure 9.** Laser scattering tomography of $Si_{1-x}Ge_x$ alloy with x = 0.50, which reveals the precipitation and the dislocation lines.

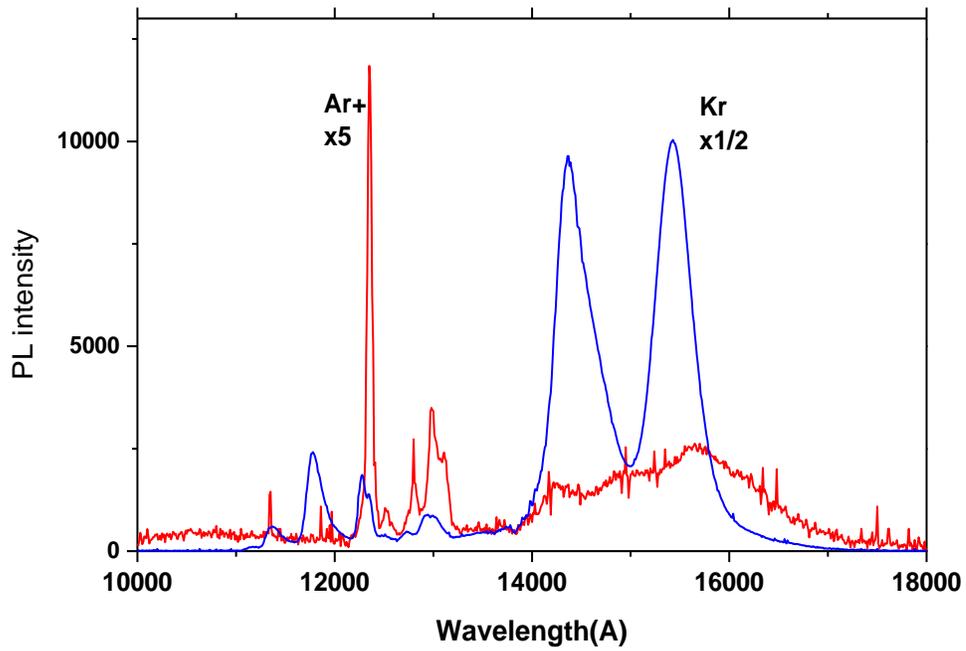

**Figure 10.** 4.2 K Ar$^+$ laser 514.5 nm line excited D-band luminescence as compared to Kr excited PL of the same sample of x = 0.50.

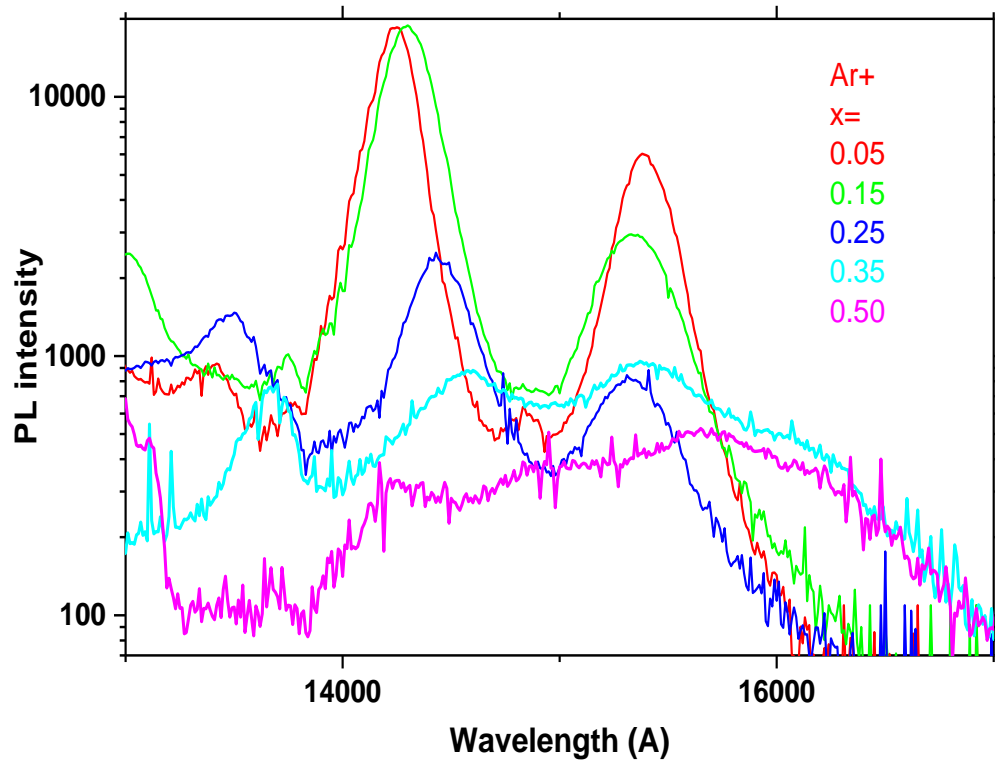

**Figure 11.** Composition dependent deep level PL emissions as excited using Ar+ laser at 4.2K.